\documentclass[11pt,a4paper]{article}
\usepackage[hyperref]{nlp4MusA}
\usepackage{times}
\usepackage{latexsym}
\usepackage{graphicx}
\usepackage{subfigure}
\usepackage{stfloats}
\usepackage{caption}
\usepackage{booktabs}

\usepackage{url}

\aclfinalcopy 

\title{Diverse Melody Generation from Chinese Lyrics via Mutual Information Maximization}

\author{Ruibin Yuan \thanks{ Email: \texttt{yuanruibin@bupt.edu.cn}} \\\And
  Ge Zhang\thanks{ Email: \texttt{zhangysk123@gmail.com}}, 
  Anqiao Yang \thanks{ Email: \texttt{yaq\_sice@bupt.edu.cn}} \\\And
  Xinyue Zhang \thanks{ Email: \texttt{zhangxinyueBUPT@outlook.com}} \\
  }

\date{2020.7}

\begin{document}
\maketitle
\begin{abstract}
In this paper, we propose to adapt the method of mutual information maximization into the task of Chinese lyrics conditioned melody generation to improve the generation quality and diversity. We employ scheduled sampling and force decoding techniques to improve the alignment between lyrics and melodies. With our method, which we called Diverse Melody Generation (DMG), a sequence-to-sequence model learns to generate diverse melodies heavily depending on the input style ids, while keeping the tonality and improving the alignment. The experimental results of subjective tests show that DMG can generate more pleasing and coherent tunes than baseline methods.

\end{abstract}

\section{Introduction}
The automatic composition has been a popular research field in the cross-discipline of music art and computer science. Despite the well-explored ones, the task of melody generation conditioned on lyrics remains a relatively new topic. It is also a marketable task since it can enable people to compose melody from texts without knowing music theory. 

Only a handful of works are focusing on applying deep-learning techniques into this task. A sequence-to-sequence framework (seq2seq) with additional context melody encoder and alignment decoder was proposed for this task~\cite{bao:2019}, using a customized closed source dataset for training. A standard seq2seq , iComposer~\cite{lee:2019}, was proposed to generate pitches and duration separately from Chinese lyrics, along with a released dataset. The latest work focusing on this task utilized adversarial learning technique~\cite{yu:2019}, and released a large scale English lyrics-melody pairs dataset. 

Inspired by previous work on bridging the gap between training and inference~\cite{zhang:2019}, and the success of the mutual information concept applying to the deep-learning~\cite{yang:2018,hjelm:2018}, 
in this paper: \textbf{(1)} We propose a novel melody generation deep-learning model conditioned on Chinese lyrics, called \emph{Diverse Melody Generation} (DMG), which incorporates the mutual information maximization technique (MIM) in the seq2seq framework to generate diverse melodies according to the input style vector. \textbf{(2)} Scheduled sampling and force decoding techniques are innovatively applied to handle the alignment between lyrics and music representations. \textbf{(3)} The subjective test shows that our proposed method can generate more pleasing and coherent tunes than the baseline methods, along with diversity and tonality.

\section{Methodology}

\subsection{DMG: Seq2Seq with MIM}
We model the task of Chinese lyrics conditioned melody generation as a type of neural machine translation task, therefore leverage a classical seq2seq with attention mechanism as underlying architecture. Rather than strictly translating input lyrics into ground-truth melodies, we hope our model can also take a style id  where $\emph{Id} \in 1,2,..., \emph{K}$
as input to generate diverse distribution of output sequences according to the style id. Since mutual information quantifies the mutual dependence between the two random variables, by concatenating a style id with the encoded context vector $\emph{v}$ and maximizing the mutual information between the output sequences and the input id, we can generate melodies that heavily depend on the input id.

Given a source sequence of $\emph{X} = \{x_{1},x_{2},...x_{T}\}$ and target sequence of
$\emph{Y} = \{y_{1},y_{2},...y_{T}\}$ where $x_{i},y_{i}$ for $\emph{i,j}=1,2,...,T$
are characters and music representations, and \emph{T} is the total number of characters in the sentence, the generating process of the seq2seq can be interpreted as conditional probability estimation:
\begin{equation}
\emph{p}(Y|X) = \prod\limits_{j=1}^{T}p(y_{t}|v,y_{1},...,y_{j-1})
\end{equation}
where \emph{v} is the fixed-dimensional representation of the source sequence. In this paper, we use bidirectional LSTM unit as the encoder of the seq2seq to project $e_{x}$,
the embedding sequence of \emph{X}, into \emph{v} a fixed-length context vector. Then we use a unidirectional LSTM unit with attention mechanism~\cite{bahdanau:2014}
as the decoder of the seq2seq to project \emph{v} into 
$P_{j}$ for $j = 1,2,...,T$, the probability distribution over decode vocabulary.

To maximize the mutual information between \emph{Id}
and \emph{Y}, following previous work on poetry generation~\cite{yang:2018} a lower bound of the mutual information $I(Id;Y)$ is computed by:
\begin{equation}
\sum_{k=1}^K p(Id=k)\int\limits_Y logQ(Id = k|Y)dY
\end{equation}
where \emph{Q} is a posterior distribution estimator, assuming that the input \emph{Id} is uniformly distributed and $p(Id=k)=\frac{1}{K}$ for
$k=1,2,...,K$ where $K$ is the total number of styles.

In previous work,Yang \shortcite{yang:2018}
leverages a soft-max normalized linear projection function as
\emph{Q} which takes the time average expected embedding of
\emph{Y} as input and outputs estimated probability of each style. However, this will lead to loss of temporal information, and encourage the model to pay attention to the frequently used special words, which is a critical problem in music generation as we will discuss in the session 3.2. To adapt the posterior distribution estimator for music sequence, we leverage a soft-max normalized $GRU$ as \emph{Q}
that processes expected embedding separately and sequentially to capture the temporal information. The lower bound can be maximized by adding it into the training objective as a regularization term. 

To calculate the integration in Eq. 2, expected character embedding  $e_{j}^{expected}$ is used to 
to approximate the probabilistic space of the $j$-th 
output character, which is computed by:
\begin{equation}
e_{j}^{expected} = P_{j}W_{embed}
\end{equation}
where $W_{embed}$ is the embedding matrix of the decoder. It enables the model to look at all the possible words at once, avoid searching through all the possible sequences, and avoid the non-differentiable decoding process. Keep feeding the $e_{j-1}^{expected}$ into the decoder with attention mechanism for $j=1,2,...,T$
we can obtain expected sequence $e_{seq}^{expected}$
, which can be fed into $GRU$ to compute the posterior probability distribution. Note that $e_{0}^{expected}$
is the embedding of tag “\_START\_”. The training objective can be written as:
\begin{equation}
  \begin{array}{l}Train(X,Y) = (1-\lambda)\sum\limits_{j=1}^{T} log p(y_{i}|y_{1},...,
y_{i-1},X) \\
    + \frac{\lambda}{K} \sum\limits_{k=1}^K log\_softmax(GRU(e_{seq}^{expected})[k])
  \end{array}
\end{equation}
where \emph{[k]} indicates the \emph{k}-th dimension of the
\emph{GRU} output score and $\lambda$ is a balance factor. Note that the regularization term can be seen as a classification loss to determine whether the output sequences incorporate the information from the style id's information. The mutual information maximization part of DMG is shown in Fig. 1.
\begin{figure}[h]
  \centering
  \includegraphics[width=3in,height=1.5in]{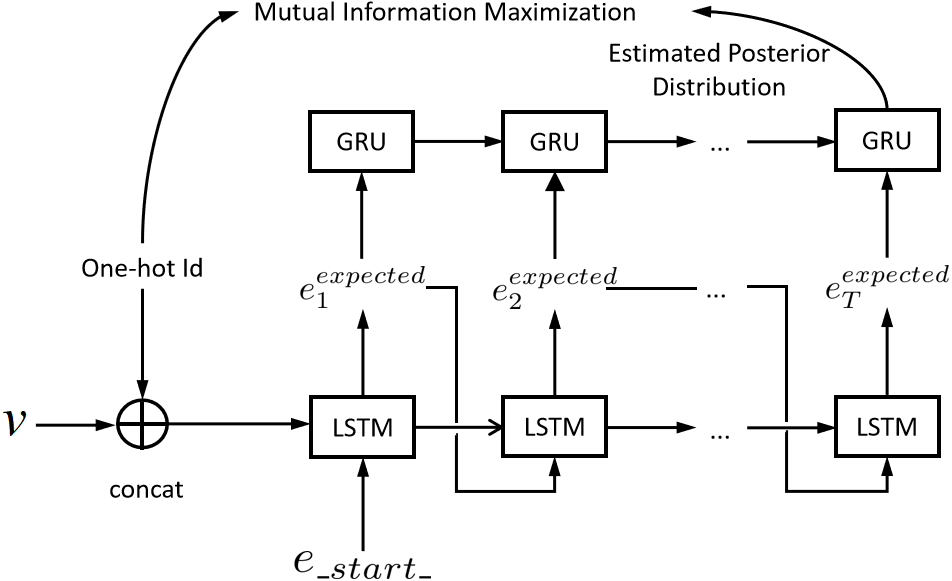}
  \caption{The MIM part of DMG.}
\end{figure}

\subsection{Scheduled Sampling and Force Decoding}
Datasets focusing on our task often suffer from the poor alignment of duration, while pitches’ alignment is relatively better. To prevent the duration's poor alignment from affecting the pitch generation result, we separately train the duration generation network and the pitch generation network. To better align the separately generated representations to the lyrics, we concatenate the corresponding representations of one character with “$,$”
as a single music character to create one-to-one alignment. Also, to teach the model to learn the musical and semantic relationship between neighboring lines, we process the two neighboring lines simultaneously and use “$|$” as a tag to indicate the boundary of different lines. However, if we only use teacher forcing during the training stage, the model does not learn the fact that music representations and input lyrics must match in number, and thus usually predicts “\_EOS\_” and “$|$” too soon or too late. To address this, we use scheduled sampling and force decoding together to teach our model to recover from its alignment mistakes and adapt to the inference stage, rather than following previous works to add additional network structures or treat it as a sequence labeling problem. 

A sampling schedule is employed to randomly select the ground truth representation $y_{j-1}^{truth}$ or the predicted representation $y_{j-1}^{pred}$ as
$y_{j-1}$. The probability of selecting the ground truth representations is \emph{P}, which is computed by:
\begin{equation}
p=\frac{\mu}{\mu + exp(\frac{epoch}{\mu})}
\end{equation}
where $\mu$ is a hyper-parameter to control the decaying rate of \emph{p},
and \emph{epoch} is the index of training epochs (starting from 0). The function is strictly monotone decreasing.

If at \emph{j}-th step where the corresponding character $x_{j}$ is not “$|$” or “\_EOS\_”,
and the top first output representation of output distribution $P_{j}$ is "$|$" or “\_EOS\_”,
we select the top second representation in $P_{j}$ as the $y_{j}^{pred}$  to force the model not to end current line too soon. Also, if at \emph{j}-th step where the corresponding character $x_{j}$ is "$|$" or “\_EOS\_”, we simply select "$|$" or “\_EOS\_” as the
$y_{j}^{pred}$. As training progresses, \emph{p} drops and the model has to deal with the error accumulation and the alignment mistakes itself, similar to the inference stage.

\section{Experiments}

\subsection{Dataset and Preprocessing}
The iComposer dataset~\cite{lee:2019} is used in this paper. We entrust people in the music industry to proofread the original data and the automatically estimated key annotations. We manage to proofread most of the key and part of the pitches annotations while leaving the duration annotations “in the wild”. Then each song is normalized into C major or A minor and the lyrics are converted into Chinese phonetic alphabets, Pinyin, as input to narrow the input vocab space, emphasize the intonation while keeping semantic to a certain extent. The dataset is split by 8:1:1 as the training set, validation set and test set. Note that the key of a song can shift as the song progresses, we only consider the key at the beginning of the song.

\subsection{Evaluation}
We consider DMG with linear projection posterior probability estimator (DMG(LP)) and iComposer
~\cite{lee:2019} as our baseline methods. To distinct DMG from DMG(LP), in this session, we call DMG as DMG(GRU). We train DMG(GRU) and baselines to epoch 100 and evaluate them on the test set. Note that the original iComposer only takes one line of characters as input instead of Pinyin, and does not normalize the input melodies into the same tune, which will lead to a significant downgrade in the tonality. For the sake of fairness, we customize a special normalized mirrored dataset for it according to the original setting, where the way we split the dataset remains the same. 

\begin{figure*}[t]
    \centering
    \subfigure[Pitch Distribution of DMG(GRU)]{
       \begin{minipage}{7cm}
       \centering
       \includegraphics[width=2.5in,height=1.5in]{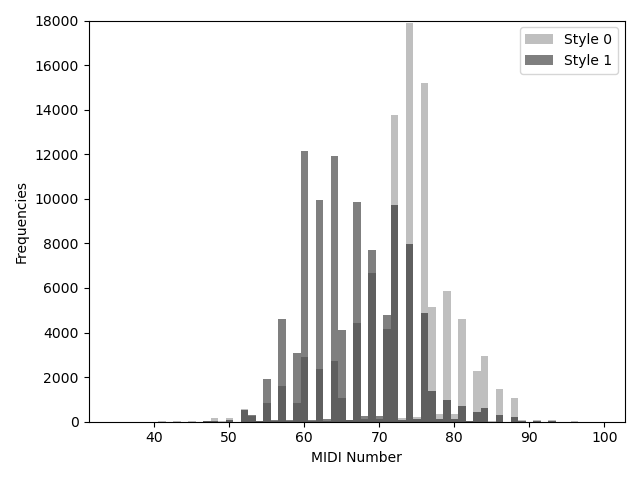}
       \end{minipage}
    }
    \subfigure[Pitch Distribution of DMG(LP)]{
       \begin{minipage}{7cm}
       \centering
       \includegraphics[width=2.5in,height=1.5in]{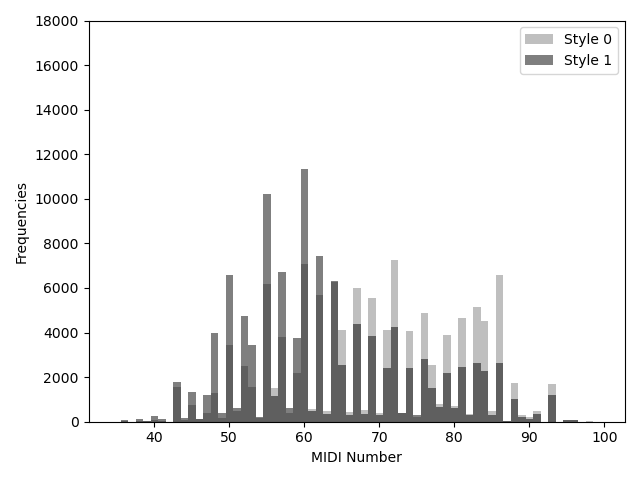}
       \end{minipage}
    }
    \caption{Generated pitch distribution on the test set when \emph{K}=2. The darkest color indicates the overlapped part.}
\end{figure*}

Next, we compare the generation results of DMG(GRU) and DMG(LP) to demonstrate the improvement after replacing the linear projection function with a \emph{GRU}. The diverse distribution of generated pitch shown in Fig. 3 indicates the regularization term’s effectiveness. However, the generated pitches’ distribution of DMG(LP) is dispersed, with a higher probability of deviating from the key, which leads to unsatisfying results demonstrated in Fig. 4. This is due to the time average pooling of expected embedding, which extracts each sequence’s topic by discovering frequently used special notes. However, after normalization, most of the songs stay in the scale of  "\emph{C,D,E,F,G,A,B}".
Therefore, none of the notes are special, except for some songs employ modulation or special scales. Hence, the embedding vector after time average pooling can only indicate these music techniques are employed into a song. According to our observation, DMG(LP) fails to properly utilize these music techniques, but only inserts random notes deviating from the key into the generated sequence. By replacing the time average pooling and linear projection function with a  \emph{GRU} which processes the expected embedding sequentially, the model can take the temporal information of the sequence into account, without paying too much attention to the frequency of special notes. As a result, generating result’s deviation from the key is reduced while the hidden space is disentangled according to the input style vectors.

\begin{figure}[h]
  \centering
  \includegraphics[width=2.5in,height=1.5in]{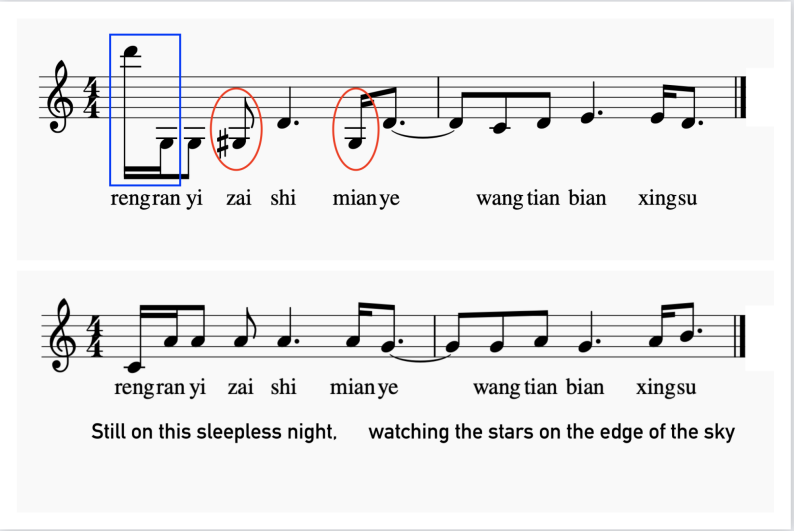}
  \caption{Sample lines from the song \emph{Half Moon Serenade}, 
  where top to bottom are generated results of DMG(LP) and DMG(GRU). The box indicates an abrupt shift in octave and 
the circles indicate the notes deviate from the key of C.
}
\end{figure}

Then we conduct a subjective test by employing five human participants with more than ten years of music training to judge the generation quality. We randomly select 20 segments of 2 neighboring lines of lyrics as encoder inputs and generate corresponding melodies using each model, mixing in original melodies composed by human. For fully automatic generation, style ids are randomly selected. Note that iComposer without force decoding and scheduled sampling cannot align the generated melodies perfectly with the lyrics, we simply duplicate or cut off the generation result to fit in the length of the lyrics. The human participants were asked to rate the randomly ordered melodies with corresponding lyrics from 0 to 5 (bad to good) in terms of the following standards: \textbf{T}onality (\emph{does the melody stick to the key?}), \textbf{C}oherency (\emph{is the melody smooth?}), \textbf{M}elodiousness (\emph{is the melody pleasing?}), \textbf{F}itness (\emph{does the melody fit the lyrics?}).

\begin{table}[htbp] 
 \centering
 \begin{tabular}{ccccc} 
  \toprule 
  Model & T & C &  M & F\\ 
  \midrule 
  iComposer & 3.31 & 2.15 & 2.58 & 2.36 \\
  DMG(LP) & 2.93 & 3.17 & 3.29 & 2.78 \\
  DMG(GRU) & 4.78 & 3.93 & 3.61 & 3.12 \\
  Human & 5.00 & 4.82 & 4.45 & 3.94 \\
  \bottomrule 
 \end{tabular} 
 \caption{Average scores of subjective test results.} 
\end{table}

Table 1 shows the human evaluation results. According to the result, DMG outperforms the baselines in all metrics. On the “Tonality” metrics, DMG(GRU) scores close to human composer, while DMG(LP) scores lower than iComposer, which indicates that using time average pooling will harm the tonality, replace it with \emph{GRU} can significantly enhance the result. The result of “Coherency” metrics suggests that by using two lines as one input, the melodies become smoother as the model learns the relationship between neighboring lines. The reduction of the abrupt shifts and enhancement of the tonality also contribute to the coherency. In terms of the “Melodiousness” metrics, we can see that using force decoding and scheduled sampling to improve the alignment can generate more pleasing melodies instead of manually cutting off or duplicating the generation results. Also, using Pinyin as input can force the model to pay more attention to the musical ideas behind the melodies instead of simply translating a word to note. The result of the “Fitness” metrics indicates that even though converting words into Pinyin will lose part of the semantic information, the model still manages to generate melodies that fit the lyrics to some extent.

\section{Conclusion and Future Work}
In this paper, we propose to adapt the mutual information maximization technique into the task of Chinese lyrics conditioned melody generation under the standard attention based sequence-to-sequence framework to improve the generation diversity and quality, along with force decoding and scheduled sampling being employed to improve the alignment between lyrics and melodies. The result shows the effectiveness of our method. For future work, we plan to improve the dataset quality and generate melodies with meaningful styles.

\bibliography{nlp4MusA}
\bibliographystyle{nlp4MusA_natbib}

\end{document}